\documentclass[paper,nofootinbib] {revtex4}

\usepackage{graphicx}
\usepackage{epsfig}
\usepackage{fancyhdr}
\usepackage{empheq}
\usepackage{mathbbol}
\usepackage{wasysym}
\usepackage{pstricks}
\usepackage{color}
\usepackage{bbm}
\def\mol{D^0\bar D^{*0}}

\begin{document}

\title{More loosely bound hadron molecules at CDF?}

%

\author{C Bignamini$^{\ocircle\dag}$, B Grinstein$^{*}$, F Piccinini$^{\dag}$, AD Polosa$^\P$, V Riquer$^\S$} 
\author{C Sabelli$^{\ddag\P}$}
\affiliation{
$^\ocircle$Dipartimento di Fisica Nucleare e Teorica, Universit\`a di Pavia, via A. Bassi 6, Pavia,  I-27100, Italy
\\
$^*$CERN-PH-TH, CH-1211 Geneva 23, Switzerland\\  and  University of California, San Diego, Department of Physics, La Jolla, CA 92093-0315, USA\\
$^\dag$ INFN Pavia, Via A. Bassi 6, Pavia, I-27100, Italy\\
$^\P$INFN Roma, Piazzale A. Moro 2, Roma, I-00185, Italy\\
$^\S$Fondazione TERA,  Via Puccini, Novara, 11 -28100, Italy\\
$^\ddag$Department of Physics, Universit\`a di Roma, `La Sapienza', Piazzale A. Moro 2, Roma, I-00185, Italy}

\begin{abstract}
In a recent paper we have proposed a method to estimate the prompt production cross section 
of $X(3872)$ at the Tevatron assuming that this particle is a loosely bound molecule of a $D$ and a $\bar D^*$
meson. Under this hypothesis we find that  it is impossible to explain the high prompt production cross section 
found by CDF  at $\sigma(X(3872))\sim30\div 70$~nb as our theoretical prediction is about 
300 times smaller than the measured one. Following our work, Artoisenet and Braaten, 
have suggested that final state interactions in the $\mol$ system might be so strong to push 
the result we obtained for the  cross section up to the experimental value. Relying on their
conclusions we show that the production of another very narrow  loosely bound 
molecule, the $X_s=D_s\bar D_s^*$, could be similarly enhanced. $X_s$ should then be detectable
at CDF with a mass of $4080$~MeV and a prompt production 
cross section  of $\sigma(X_s)\sim 1\div 3$~nb. 
\end{abstract}

\maketitle

\thispagestyle{fancy}

{\bf \emph{Introduction}}.
In a recent paper~\cite{bignamini} we have proposed a method for estimating the prompt production cross section of $X(3872)$~\cite{bellex} 
at the Tevatron making the assumption that $X$ is a loosely bound molecule of $D$ and  $\bar D^*$, with a binding energy as small 
as ${\cal E}_0=-0.25\pm0.40$~MeV. The motivation for this study is that, after the Belle discovery, CDF and D0 confirmed the $X(3872)$ 
in proton-antiproton collisions~\cite{cdfx,d0x} and it  seems at odds with common intuition that  such a loosely bound molecule could be 
produced promptly ({\it i.e.} not from $B$ decay) in a high energy hadron collision environment. This was also one of the initial motivations to 
consider the possibility that the $X(3872)$ could be, instead of a molecule, a `point-like' hadron resulting from  the binding of a diquark 
and an antidiquark~\cite{4q}, following the interpretation proposed by Jaffe and Wilczek~\cite{jw} of pentaquark baryons 
(antidiquark-antidiquark-quark). 

To start let us summarize the content of~\cite{bignamini}.
Let us suppose that $X(3872)$ is an $S$-wave bound state of two $D$ mesons, namely 
a $1/\sqrt{2}(D^0\bar D^{*0}+\bar D^0 D^{*0})$ molecule (we will  use the shorthand notation $D^0\bar D^{*0}$)
~\footnote{Such a molecule has the correct $1^{++}$ quantum numbers of the $X(3872)$.}. The molecule production 
cross section will be proportional to the number of $\mol$ pairs in the event. 
Thus the $X(3872)$ prompt production cross section at the Tevatron could be written as:
\begin{eqnarray}
&&\sigma(p\bar p\to X(3872)) \sim \left| \int d^3 {\bf k} \langle X|D\bar D^{*}({\bf k})\rangle\langle D\bar D^{*}({\bf k})|p\bar p\rangle\right|^2
\simeq \left| \int_{\cal R} d^3 {\bf k} \langle X|D\bar D^{*}({\bf k})\rangle\langle D\bar D^{*}({\bf k})|p\bar p\rangle\right|^2 \nonumber\\
&&\leq \int_{\cal R} d^3 {\bf k} |\psi({\bf k})|^2 \int_{\cal R} d^3 {\bf k} | \langle D\bar D^{*}({\bf k})|p\bar p\rangle |^2
\leq\int_{\cal R} d^3 {\bf k} | \langle D\bar D^{*}({\bf k})|p\bar p\rangle |^2\sim \sigma(p\bar p\to X(3872))^{\rm max}
\label{eq:erre}
\end{eqnarray}
where ${\bf k}$ is the relative 3-momentum between the $D({\bf p_1}),D^*({\bf p_2})$ mesons.
$\psi ({\bf k})=\langle X|D\bar D^{*}({\bf k})\rangle$ is some normalized bound state wave function characterizing the $X(3872)$. ${\cal R}$ is the integration region where $\psi({\bf k})$ is significantly different from zero. 
The  matrix element $\langle D\bar D^{*}({\bf k})|p\bar p\rangle$ can be computed using standard 
matrix-element/hadronization Monte Carlo programs (MC) like Herwig~\cite{herwig} and Pythia~\cite{pythia}~\footnote{Open charm meson pairs generated with hadronization Monte Carlo 
are ordered as a function of their relative center-of-mass 3-momenta. 
If more than one $D^0\bar D^{*0}$ pair is found in the event, we select the pair having the smaller relative 3-momentum $k$. 
As a first step we select those pairs which  pass the kinematical cuts used in the 
data analysis made by the CDF collaboration.}. We require our MC tools to generate $2\to 2$ QCD  events with some loose partonic generation cuts as detailed in~\cite{bignamini}~\footnote{Configurations with one gluon recoiling from a $c\bar c$ pair, are those configuration expected to produce two collinear charm quarks and in turn collinear open charm mesons. The parton shower algorithms in Herwig and Pythia treat properly these configurations at 
low $p_\perp$ whereas they are expected to be less important at  higher $p_\perp$.}.

As for the determination of the region ${\cal R}$ in (\ref{eq:erre}) we estimate it having in mind a naive gaussian ansatz for the bound state wave function.
It is straightforward to estimate the momentum spread of the gaussian by  assuming a strong interaction 
Yukawa potential between the two $D$ mesons. Given that
the binding energy ${\cal E}_0$ is ${\cal E}_0\sim M_X - M_D- M_{D^*}= -0.25\pm 0.40$~MeV
 we find that $r_0\sim 8$~fm ($8.6\pm1.1$~fm) and
applying the (minimal) uncertainty principle relation, we get the gaussian momentum
spread  $\Delta p \sim 12$~MeV. 

Given the very small binding energy  we can estimate $k$ to be as large as $k\simeq \sqrt{2\mu (-0.25+0.40)}\simeq 17$~MeV, $\mu$ being the reduced mass of $\mol$ system, or of the order of the center of mass momentum
$k=\sqrt{\lambda(m_X^2,m_D^2,m_D^{*2})}/2m_X\simeq 27$~MeV. 
These considerations imply that 
we can restrict the integration region to  a ball ${\cal R}$ of radius~\footnote{Which corresponds to a $k_0$ of the Gaussian at $\sim 27$~MeV and a spread of $+12$~MeV.} $\simeq[0,35]$~MeV.

Keeping $k$ inside ${\cal R}$ we estimate a $\sigma(p\bar p\to X(3872))^{\rm max}$ which is about 30 times smaller than the most conservative estimate 
($\sigma\sim 3.2$~nb) of the 
minimal prompt production cross section measured at CDF~\cite{bignamini}. In the analysis proposed in~\cite{braaten} the experimental cross section is estimated  from data  to be
$\sigma\sim 30\div 70$~nb reinforcing the negative result obtained with our theoretical 
calculation that would rather be 300 times smaller than the measured one.
This fact would undoubtably put in serious trouble the molecular interpretation of $X(3872)$. 

\begin{figure}[h!]
\includegraphics[width=7truecm]{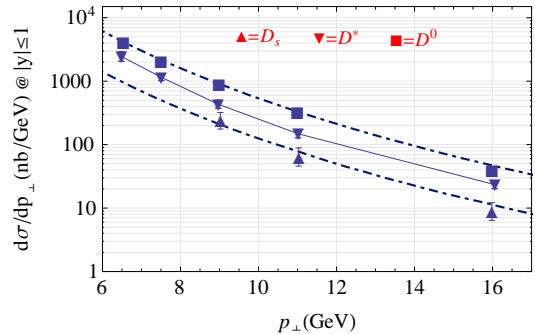}
\caption{
Differential charm cross section measured in fully hadronic charm decays using $5.8~{\rm pb}^{-1}$ at CDF~\cite{cdfdat}. 
The error bars represent the total uncertainty of the measurement. The ratio in the production of $D^0$ and $D_s$ is 4.4 whereas the ratios in the production of $D^*$ and $D_s$ is about 2.2.
The dot-dashed lines are the result of the Monte Carlo simulation done with Herwig rescaling the normalizations of the distributions by a factor $K=1.5$. This value
is in very good agreement with the $K$ factor found in~\cite{bignamini} ($K=1.8$) 
using data on $d\sigma/d\Delta\phi$, $\Delta\phi$ being the angle between $D^0$ 
and $D^{*-}$ mesons produced at CDF within some definite cuts in rapidity and transverse momentum.}
\label{data}
\end{figure}

In this note we intend to start from the main result  discussed in~\cite{braaten} where it is argued 
that the effect of final state interactions in the
$\mol$ system is such that two corrections should be made to our previous calculation: 
$i)$ the ball ${\cal R}$ should be enlarged to include momenta up to $\Lambda\sim 300$~MeV; $ii)$ a correction factor to the cross section we compute (see~(\ref{eq:erre})) should be considered so that the actual cross section $\sigma^*$ including the full effect of final state interaction is 
\begin{equation}
\sigma^*(p\bar p\to X(3872))=\sigma(k<\Lambda)\times\frac{6\pi\sqrt{2\mu |{\cal E}_0|}}{\Lambda}
\label{brat}
\end{equation}
Assuming that this is  the correct way of  discussing the $X(3872)$ production we observe that, 
besides reconciling the experimental result with the theoretical computation for the $X$, this mechanism 
should enhance the occurrence of an hypothetical new molecule, the $D_s\bar D_s^*$, which otherwise would be suppressed as one could infer by looking at data on $D_s$ production at Tevatron~\cite{cdfdat} (as shown in  Fig.~\ref{data}, $D_s$ is on average $\sim5$ times less probable than $D^0$). 
The same data are used to tune our Monte Carlo (Herwig in this calculation) 
with respect to $D_s$ production as shown in Fig.~\ref{data}.
 
The $X_s(1^{++})$ molecule should exist as a partner with strange light quarks of the $X(3872)$. One could expect it to be a more compact 
molecule with respect to the $X(3872)$ as $\eta$ particle exchange forces would be at work. This enlarges the spread $\Delta p$ 
in the relative momentum and naturally makes the ball ${\cal R}$ larger.  We will
postulate a binding energy  for $X_s$  as small as that of the $X(3872)$ and its mass is expected to be $M_{X_s}=4080$~MeV. 
$X_s$ could decay into $J/\psi \pi\pi$ with a narrow width because of its  mass and  flavor content (it
cannot decay into $K^+K^- J/\psi$ (via $\phi$) because of phase space; it cannot either
decay to $J/\psi f_0(980)$ because of quantum numbers), or in $D_sD_s\gamma$.
Using the Herwig hadronization algorithm to compute $\sigma(k<\Lambda)$  in~(\ref{brat}) we obtain
\begin{equation}
\sigma^*(p\bar p\to X_s(4080))=1\div 3~{\rm nb}
\end{equation}
where the value of $3$~nb is found pushing the $\Lambda$ value up to $600$~MeV (following some considerations on the possible values of the $\Lambda$ cutoff made in~\cite{braaten}).
We obtain definitely similar results using Pythia~\cite{pythia}.

Such numbers should put the $X_s(4080)$ molecule in the conditions to be observed at CDF. 
We would find rather surprising that no such state is found assuming that  the mechanism~(\ref{brat}) is correct thus we encourage searches of this resonance.

On the other hand we cast some doubts on the possibility that final state interactions can indeed play such a pivotal role as described in~\cite{braaten}. First of all we remind that Watson formulae~\cite{watson} used in~\cite{braaten} are valid for $S$-wave scattering, whereas a relative three-momentum $k$ of 
$300$~MeV indicates that higher partial waves should be taken into account. 

\begin{figure}
\includegraphics[width=7truecm]{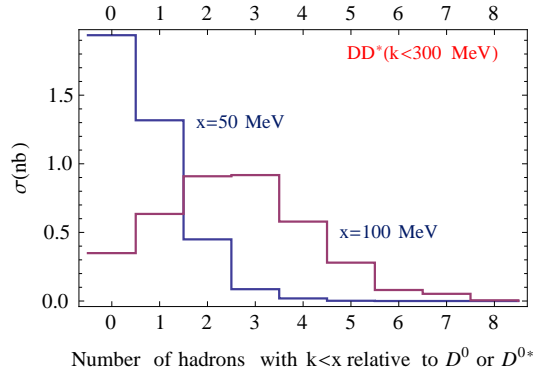}
\caption{The cross section integrated in  bins containing $n=0,1,2,...$ extra hadrons having a relative momentum $k<x$~MeV with respect the $D$ or the $D^*$ composing the $X(3872)$ molecule. Following~\cite{braaten} we assume that the molecule is formed in $S$-wave with a relative $k$ in the center of mass of $D$ and $D^*$ as large as 300 MeV.}
\label{nhad}
\end{figure}

Most importantly, we have verified in our MC simulations that as the relative momentum $k$  in the center of mass of the molecule is taken to be up to $300$~MeV, then other hadrons (on overage more than two) have a relative momentum $k<100$~MeV with the $D$ or the $D^*$ 
constituting the molecule (see Fig.~\ref{nhad}). 
On the other hand the Migdal-Watson theorem for final state interactions
requires that {\it only two} particles in the final state participate to the strong interactions causing them to rescatter. In other words the extra hadrons involved in the process do necessarily interfere in an unknown way with the mesons assumed to rescatter into an $X(3872)$. This is particularly true as one further enlarges the dimensions of the momentum ball ${\cal R}$ as required in~\cite{braaten}.

Tetraquarks with a $[cs][\bar c \bar s]$ might also occur, and one expects the lightest of this family to be a scalar at about 3930~MeV, as estimated in~\cite{drenska}. Computing the prompt production cross section is an harder task though.  This would require some specific model for the
fragmentation of partons into diquarks allowing to extract from data a ratio of the production rate of $[cs]$ and $[cq]$ diquarks. In turn this would allow, for example, to estimate the prompt production cross section of the $X_s$ under the hypothesis that the $X(3872)$ produced at CDF is a tetraquark. A simple model of parton to diquark fragmentation could be drawn along the lines discussed in~\cite{maianis} where the case of light diquarks was treated. Yet we prefer to postpone such estimate as soon as the first data on exotic hadron production will be available from LHCb and ALICE.

In this note we show that starting from the results discussed in~\cite{braaten} we should expect an enhancement in the prompt  production cross section of  an hypothetical new $X_s(4080)$ molecular loosely bound resonance constituted by a $D_s\bar D_s^*$ pair. We estimate such cross section to be between 1 and 3 nb at the Tevatron. On the other hand we cast some doubts on the applicability of the Watson theorem for final state interactions in the calculation at hand. We show that in the hadronization shower the number of hadrons in a momentum volume 
${\cal R}(k)$ tends to grow with $k$ whereas the final state interactions formulae used in~\cite{braaten} (see~\cite{watson}) should involve only two hadrons at a time. 

\begin{acknowledgments}
We wish to thank Marco Rescigno for his indispensable hints on CDF data.
The work of one of us (B.G.) is supported in part by the US Department of Energy under contract DE-FG03-97ER40546.

\end{acknowledgments}

\bigskip 

\end{document}